\documentclass[twocolumn,showpacs,preprintnumbers,amsmath,amssymb]{revtex4}
\usepackage{dcolumn}
\usepackage{bm}

\begin {document}

\preprint{quant-ph/0606040}

\title {On the additivity conjecture for the Weyl channels being covariant with respect
to the maximum commutative group of unitaries}

\author {Grigori G. Amosov}

\email {gramos@mail.ru}

\affiliation {Department of Higher Mathematics\\
Moscow Institute of Physics and Technology\\ Dolgoprudny
141700\\RUSSIA}

\date{\today}

\begin {abstract}
Some new examples of quantum channels for which the infimum of
the output entropy is additive under taking a tensor product of
channels are given.
\end {abstract}


\maketitle

A linear trace-preserving map $\Phi $ on the set of states
(positive unit-trace operators) $\sigma (H)$ in a Hilbert space
$H$ is said to be a quantum channel if $\Phi ^{*}$ is completely
positive (\cite {Hol}). The channel $\Phi $ is called
bistochastic if $\Phi (\frac {1}{d}I_{H})=\frac {1}{d}I_{H}$.
Here and in the following we denote by $d$ and $I_{H}$ the
dimension of $H,\ dimH=d<+\infty ,$ and the identity operator in
$H$, respectively. Fix the basis $|f_{j}>\equiv |j>,\ 0\le j\le
d-1,$ of the Hilbert space $H$. We shall consider a special
subclass of the bistochastic Weyl channels (\cite {Amo, Amo1,
Ruskai, Fukuda, Cerf}) defined by the formula (\cite {Amo1})
\begin {equation}\label {Weyl}
\Phi (x)=(1-(d-1)(r+dp))x+r\sum \limits
_{m=1}^{d-1}W_{m,0}xW_{m,0}^{*}
\end {equation}
$$
+p\sum \limits _{m=0}^{d-1}\sum \limits
_{n=1}^{d-1}W_{m,n}xW_{m,n}^{*},
$$
$x\in \sigma (H)$, where $r,p\ge 0,\ (d-1)(r+dp)=1$ and the Weyl
operators $W_{m,n}$ are determined as follows
$$
W_{m,n}=\sum \limits _{k=0}^{d-1}e^{\frac {2\pi i}{d}kn}|k+m\ mod\
d
><k|,
$$
$0\le m,n\le d-1$.

 Consider the maximum commutative group ${\mathcal U}_{d}$
consisting of unitary operators
$$
U=\sum \limits _{j=0}^{d-1}e^{i\phi _{j}}|e_{j}><e_{j}|,
$$
where the orthonormal basis $(e_{j})$ is defined by the formula
$$
|e_{j}>=\frac {1}{\sqrt d}\sum \limits _{k=0}^{d-1}e^{\frac {2\pi
i}{d}jk}|k>,\ 0\le j\le d-1,
$$
$\phi _{j}\in {\mathbb R},\ 0\le j\le d-1$. Notice that
$$
<f_{k}|e_{j}>=\frac {1}{\sqrt d}e^{\frac {2\pi i}{d}jk},\ 0\le
j,k\le d-1,
$$
It implies that
\begin {equation}\label {MUB}
|<f_{k}|e_{j}>|=\frac {1}{\sqrt d}
\end {equation}
The bases $(f_{j})$ and $(e_{j})$ satisfying the property (\ref
{MUB}) are said to be mutually unbiased (\cite {Ivan}). It is
straightforward to check that
\begin {equation}\label {shift}
W_{0,n}|e_{j}><e_{j}|W_{0,n}^{*}=|e_{j+n\ mod\ d}><e_{j+n\ mod\
d}|,
\end {equation}
$0\le j,n\le d-1$.

It was shown in \cite {Amo1} that the Weyl channels (\ref {Weyl})
are covariant with respect to the group ${\mathcal U}_{d}$ such
that
$$
\Phi (UxU^{*})=U\Phi (x)U^{*},\ x\in \sigma (H),\ U\in {\mathcal
U}_{d}.
$$

The infimum of the output entropy of a quantum channel $\Phi $ is
defined by the formula
$$
\chi (\Phi)=\inf \limits _{x\in \sigma (H)}S(\Phi (x)),
$$
where $S(x)=-Tr(xlog(x))$ is the von Neumann entropy of the state
$x\in \sigma (H)$. The additivity conjecture for the quantity
$\chi (\Phi)$ states (\cite {AHW})
$$
\chi (\Phi\otimes \Psi)=\chi (\Phi)+\chi (\Psi)
$$
for an arbitrary quantum channel $\Psi $.

{\bf Example 1.} Put $r=p=\frac {q}{d^{2}},\ 0\le q\le 1$, then it
can be shown (\cite {Amo, Amo1, Ruskai}) that (\ref {Weyl}) is the
quantum depolarizing channel,
\begin {equation}\label {dep}
\Phi _{dep}(x)=(1-q)x+\frac {q}{d}I_{H},\ x\in \sigma (H),
\end {equation}
$$
\chi (\Phi _{dep})=-(1-\frac {d-1}{d}q)\log (1-\frac
{d-1}{d}q)-(d-1)\frac {q}{d}\log \frac {q}{d}.
$$
$\Box $

{\bf Example 2.} Put $r=\frac {1}{d}(1-\frac {d-1}{d}q),\ p=\frac
{q}{d^{2}},\ 0\le q\le \frac {d}{d-1}$, then (\ref {Weyl}) is
q-c-channel (\cite {Hol2}). Indeed, under the conditions given
above the channel $\Phi \equiv \Phi _{qc}$ can be represented as
follows
$$
\Phi _{qc}(x)=(1-\frac {d-1}{d}q)E(x)+\frac {q}{d}\sum \limits
_{n=1}^{d-1}W_{0,n}E(x)W_{0,n},
$$
where
$$
E(x)=\frac {1}{d}\sum \limits _{m=0}^{d-1}W_{m,0}xW_{m,0}^{*},
$$
$x\in \sigma (H)$ is a conditional expectation on the algebra
generated by the projections $|e_{j}><e_{j}|,\ 0\le j\le d-1$.
Taking into account (\ref {shift}) we get
\begin {equation}\label {qc}
\Phi _{qc}(x)=\sum \limits _{j=0}^{d-1}Tr(|e_{j}><e_{j}|x)x_{j},\
x\in \sigma (H),
\end {equation}
where
$$
x_{j}=(1-\frac {d-1}{d}q)|e_{j}><e_{j}|+
$$
$$
\frac {q}{d}\sum \limits _{k=1}^{d-1}|e_{j+k\ mod\ d}><e_{j+k\
mod\ d}|,
$$
$0\le j\le d-1$,
$$
\chi (\Phi _{qc})=-(1-\frac {d-1}{d}q)\log (1-\frac
{d-1}{d}q)-(d-1)\frac {q}{d}\log \frac {q}{d}.
$$

$\Box $

In the present paper our goal is to prove the following theorem.
We shall use the approach introduced in (\cite {Amo, Amo1}).

{\bf Theorem.}{\it Suppose that $d$ is a prime number and $p\le
r\le \frac {1}{d}(1-d(d-1)p)$. Then, for the channel (\ref
{Weyl}) there exist $d$ orthonormal bases
$(h_{j}^{s})_{j=0}^{d-1},\ 0\le s\le d-1,$ in $H$ such that
$$
S((\Phi \otimes \Psi)(x))\ge \chi (\Phi)+\frac {1}{d^{2}}\sum
\limits _{s=0}^{d-1}\sum \limits _{j=0}^{d-1}S(\Psi (x_{j}^{s})),
$$
$x\in \sigma (H\otimes K),$ where
$x_{j}^{s}=dTr_{H}(|h_{j}^{s}><h_{j}^{s}|x)\in \sigma (K),\ 0\le
j\le d-1,$ and $\Psi $ is an arbitrary quantum channel in a
Hilbert space $K$. }

The proof of the Theorem is based upon Theorem 2 from \cite
{Amo1}. We shall formulate it here for the convenience.

{\bf Theorem 2 (\cite {Amo1}).}{\it Let $\Phi
(\rho)=(1-p)\rho+\frac {p}{d}I_{H},\ \rho\in \sigma (H),\ 0\le
p\le \frac {d^{2}}{d^{2}-1},$ be the quantum depolarizing channel
in the Hilbert space $H$ of the prime dimension $d$. Then, there
exist $d$ orthonormal bases $\{f_{j}^{s},\ 0\le s,j\le d-1\}$ in
$H$ such that
\begin {equation}\label {XJ}
S((\Phi \otimes Id)(x))\ge -(1-\frac {d-1}{d}p)\log (1-\frac
{d-1}{d}p)-
\end {equation}
$$
\frac {d-1}{d}p\log \frac {p}{d}+ \frac {1}{d^{2}}\sum \limits
_{j=0}^{d-1}\sum \limits _{s=0}^{d-1}S(x_{j}^{s}),
$$
where $x\in \sigma (H\otimes K),\
x_{j}^{s}=dTr_{H}((|e_{j}^{s}><e_{j}^{s}|\otimes I_{K})x)\in
\sigma (K),\ 0\le j,s\le d-1$. }

Proof.

It follows from the condition $p\le r\le \frac {1}{d}(1-d(d-1)p)$
that there exists a number $\lambda ,\ 0\le \lambda \le 1,$ such
that $r=\lambda p+(1-\lambda)\frac {1}{d}(1-d(d-1)p)$. Hence the
channel (\ref {Weyl}) can be represented as a convex linear
combination of the following form
$$
\Phi =\lambda \Phi _{dep}+(1-\lambda)\Phi _{qc},
$$
where the channels $\Phi _{dep}$ and $\Phi _{qc}$ are defined by
the formulae (\ref {dep}) and (\ref {qc}), respectively.

Let us define the phase damping channel $\Xi $ by the formula
(\cite {Amo, Amo1})
$$
\Xi (x)=\frac {1+(d-1)\lambda}{d}x+\frac {1-\lambda}{d}\sum
\limits _{m=1}^{d-1}W_{m,0}xW_{m,0}^{*},\ x\in \sigma (H).
$$
Then,
$$
\Phi =\Xi \circ \Phi _{dep}.
$$
The non-decreasing property of the von Neumann entropy gives us
the estimation
\begin {equation}\label {F}
S((\Phi\otimes \Psi)(x))\ge S((\Phi _{dep}\otimes
\Psi)(x))=S((\Phi _{dep}\otimes Id)(\tilde x)),
\end {equation}
where $\tilde x=(Id\otimes \Psi)(x)$. Applying Theorem 2 to the
right hand side of (\ref {F}) we obtain the result.

$\Box $


\begin {thebibliography}{99}

\bibitem {Amo} Amosov G.G. Remark on the additivity conjecture for
the depolarizing quantum channel. Probl. Inf. Transm. 42 (2006)
3-11. e-print quant-ph/0408004.

\bibitem {Amo1} Amosov G.G. On the Weyl  channels being covariant
with respect to the maximum commutative group of unitaries.
e-print quant-ph/0605177.


\bibitem {AHW} Amosov G.G., Holevo A.S., Werner R.F. On some additivity problems in
quantum information theory.  Probl. Inf. Transm. 2000. V. 36. N 4.
P. 24-34; e-print quant-ph/0003002.

\bibitem {Ruskai} Datta N, Ruskai M.B. Maximal output purity and capacity for asymmetric unital qudit channels
J. Physics A: Mathematical and General 38 (2005) 9785-9802.
e-print quant-ph/0505048.

\bibitem {Fukuda} Fukuda M., Holevo A.S. On Weyl-covariant
channels. e-print quant-ph/0510148.


\bibitem {Hol} Holevo A.S. On the mathematical theory of quantum communication
channels. Probl. Inf. Transm. 8 (1972) 62 - 71.

\bibitem {Hol1} Holevo A.S. Some estimates for the amount of information
transmittable by a quantum communications channel. (Russian)
Probl. Inf. Transm. 9 (1973) 3 - 11.

\bibitem {Hol2} Holevo A.S. Quantum coding theorems. Russ.
Math. Surveys 53 (1998) 1295-1331; e-print quant-ph/9808023.

\bibitem {Ivan} Ivanovich I.D. Geometrical description of quantum state
determination. J. Physics A 14 (1981) 3241-3245.


\bibitem {Cerf} Karpov E., Daems D., Cerf N.J. Entanglement
enhanced classical capacity of quantum communication channels
with correlated noise in arbitrary dimensions. e-print
quant-ph/0603286.

\end {thebibliography}
\end {document}